\documentclass[preprint, showpacs, 
eqsecnum,aps,superscriptaddress, 
]{revtex4}

\newcommand{\be}{\begin{equation}}
\newcommand{\ee}{\end{equation}}
\newcommand{\bea}{\begin{eqnarray}}
\newcommand{\eea}{\end{eqnarray}}

\newcommand{\dv}{\partial}

\begin{document}
\title{Stochastic Perturbations in Vortex Tube Dynamics}
\author{L. Moriconi}
\affiliation{Instituto de F\'\i sica, Universidade Federal do Rio de Janeiro, \\
C.P. 68528, 21945-970, Rio de Janeiro, RJ, Brazil}
\author{F.A.S. Nobre}
\affiliation{Centro Brasileiro de Pesquisas F\'\i sicas, \\
Rua Xavier Sigaud, 150, Rio de Janeiro, RJ -- 22290-180, Brazil}
\affiliation{Universidade Regional do Cariri, \\
Rua Cel. Ant\^onio Luis, 1160, Crato, CE -- 63100-000, Brazil}
\begin{abstract}
A dual lattice vortex formulation of homogeneous turbulence is developed, within the Martin-Siggia-Rose field theoretical approach. It consists of a generalization of the usual dipole version of the Navier-Stokes equations, known to hold in the limit of vanishing external forcing. We investigate, as a straightforward application of our formalism, the dynamics of closed vortex tubes, randomly stirred at large length scales by gaussian stochastic forces. We find that besides the usual self-induced propagation, the vortex tube evolution may be effectively modeled through the introduction of an additional white-noise correlated velocity field background. The resulting phenomenological picture is closely related to observations previously reported from a wavelet decomposition analysis of turbulent flow configurations.
\end{abstract}
\maketitle
\section{Introduction}
A considerable progress has been achieved along the last two decades concerning the kinematics of turbulent coherent structures, a fact intimately associated to the improving performance of computer and experimental resources \cite{sreenv}. However, the relevant dynamical properties of the evolution and interaction of the energy-containing eddies -- believed by many to comprise the key for a fundamental understanding of intermittency and other turbulence characteristics -- are still essentially unknown. As a concrete illustration of the present theoretical limitations, it is worth recalling the difficulties faced in the study of wall turbulence. Even though the main flow patterns have been identified in that situation \cite{robin, land}, there is, for instance, no solid theoretical foundation for the logarithmic law of the wall. 

An ideal arena for the investigation of the dynamical and kinematical issues is provided by homogeneous isotropic turbulence. Direct numerical simulations have showed clearly that at moderately high Reynolds numbers the flow is dominated by long-lived vortex tubes with small cross-sectional dimensions (defined around the Kolmogorov dissipation length) and sizes extending up to the integral scale \cite{siggia,kerr,she,hoso,vinc}. It is also known, as set on a firm ground by Farge et al. \cite{farge}, through wavelet decomposition methods, that most of the turbulent kinetic energy is carried by the vortex tubes, which are surrounded on their turn by a background incoherent flow. 

Several analytical studies have addressed over the years the picture of homogenous turbulent flows in terms of vortex tubes, either from the dynamical or kinematical viewpoints (see Ref. \cite{pullin-saffman} for a comprehensive review). Among the former, a growing attention has been devoted to Lundgren's model \cite{lundgren}, based on the evolution of strained spiral vortices, which are transformed into tube-like structures and are probably generated in real flows through shear layer instabilities \cite{vinc,cuypers}. In contrast, in the kinematical approach, the dynamical details are bypassed, and an effective account of the statistical stationary regime of the ``vortex tube gas" is attempted, as in the works of Chorin \cite{chorin}, where a connection with standard polymer statistical mechanics is drawn, and Hatakeyama and Kambe \cite{hata}, whose focus relies on the properties of flow configurations related to multifractal distributions of vortex filaments (modeled as Burgers vortices). 

Our initial aim in this paper is to establish, in Sec. II, an alternative formulation of the turbulence problem, incorporating into the usual stochastic approach \cite{wyld} the physical insight suggested from experimental and numerical investigations, which, as commented above, place vorticity coherent structures on a central stage. More specifically, we will implement, with the help of the Martin-Siggia-Rose functional formalism \cite{msr}, an exact statistical lattice vortex description of the flow's dynamics \cite{williams,tagu,chorin}, which contains, as a special case, the known dipole form of the Navier-Stokes equations. Next, in Sec. III, having in mind modeling matters (and, thus, non rigorous arguments), we use the lattice vortex formalism just developed to advance a phenomenological scheme describing the evolution of vortex tubes forced at large scales by stochastic forces. In particular, we also consider the effective force-force correlation function employed in the renormalization group analysis of turbulence \cite{fns,d-m,y-o,smith}, decaying in Fourier space as $k^{-3}$. In Sec. IV, we find that the stochastic perturbations due to the random external forcing may be effectively interpreted as resulting from the vortex tube advection by a white-noise correlated velocity background flow. We determine the one-dimensional energy spectrum $E(k) \sim k^2$ of the background flow, including its dependence upon the energy transfer rate, and the integral and viscous scales as well. It is interesting to note that a ``thermal-like" energy spectrum, superimposed to the Kolmogorov one, was indeed observed in the numerical wavelet analysis of turbulent configurations performed in Ref. \cite{farge}. To conclude, in Sec. V, we summarize and discuss our main results.

\section{Dual Lattice Vortex Formulation}
As largely known, a systematic approach to the statistical description of homogenous isotropic turbulence, which is concerned with the flow's small scale properties, is yield by the stochastic generalization of the Navier-Stokes equations \cite{wyld},
\bea
&&\dv_t v_\alpha + v_\beta \dv_\beta v_\alpha = \nu \dv^2 v_\alpha
-\dv_\alpha P + f_\alpha \ , \ \nonumber \\
&&\dv_\alpha v_\alpha = 0 \ . \ \label{sns-eqs}
\eea
Above, $f_\alpha = f_\alpha (\vec x,t)$ denotes a gaussian random force, defined at some large length scale $L$, with vanishing expectation value and the two-point correlation function
\be
\langle f_\alpha (\vec x,t) f_\beta (\vec x',t') \rangle = \delta_{\alpha \beta} F(|\vec x - \vec x'|) \delta(t-t') \ . \ \label{tpc}
\ee
It follows from Novikov's theorem \cite{novik} that energy is injected at large scales with pumping rate ${\cal{E}}=\langle f_\alpha v_\alpha \rangle = F(0) \equiv D_0$. Furthermore, according to the standard Kolmogorov phenomenology \cite{frisch}, it is conjectured that dissipation takes place around the microscopic scale given by $\eta \sim D_0^{-1/4}\nu^{3/4}$, where viscous effects become relevant. The Reynolds number, depending only on the extreme scales $L$ and $\eta$, is $R_e \sim (L/\eta)^{4/3}$. 

Let the spatial part of force-force correlator be written as
\bea
F(|\vec x - \vec x'|) &=& { {D_0 m} \over {\pi^2}}
\int d^3 \vec k { { \exp (i \vec k \cdot \vec x) } \over
{(k^2 + m^2)^2 } }  \nonumber \\
&=& D_0 \exp ( - m |\vec x - \vec x'| ) \ , \ \label{<ff>}
\eea
with $m \equiv 1/L$. An important feature of expression (\ref{<ff>}), which does not necessarily hold for other admissible choices of the force-force correlation function, is that its inverse has a simple local form. Actually, we get, from (\ref{<ff>}), 
\be
F^{-1}(|\vec x - \vec x'|) = \frac{1}{8 \pi D_0 m} (\partial^2 -m^2)^2 \delta^3(\vec x - \vec x') \ . \
\ee
Notwithstanding the fact that the Fourier transform of $F(|\vec x - \vec x'|)$ be regarded in principle as a regularized version of Dirac's delta-function in some appropriate functional space \cite{orsz-etal}, we will discuss, later on, dynamical effects related to the alternative definition
\be
F(|\vec x - \vec x'|) = \frac{D_0}{4 \pi} \int d^3 \vec k k^{-3} \exp [i \vec k \cdot (\vec x-\vec x')]  \ , \ \label{<ffb>}
\ee
which has been a crucial ingredient in the renormalization group studies of turbulence \cite{fns,d-m,y-o,smith}. In (\ref{<ffb>}), the integration in Fourier space is bound to the region $1/L < k < 1/\eta$. Observe that in this case, the mean energy input rate per octave is fixed to $D_0 \ln 2$, and we have $F(0)= D_0 \ln(L/\eta)$.

Considering the bulk of experimental and numerical evidence that favors the picture of turbulence as a vortex tube gas, from now on our attention will be focused on the vorticity dynamics implied by the stochastic Navier-Stokes equations. It is convenient, thus, to work with the stochastic 
Helmholtz equation, straightforwardly derived from Eqs. (\ref{sns-eqs}) as
\be
\dv_t \omega_\alpha + v_\beta \dv_\beta \omega_\alpha
-\omega_\beta \dv_\beta v_\alpha = \nu \dv^2 \omega_\alpha
+  f_\alpha^\star \ , \ \label{sh-eq}
\ee
where $ f_\alpha^\star =\epsilon_{\alpha \beta \gamma} \dv_\beta f_\gamma$
and $\omega_\alpha = \epsilon_{\alpha \beta \gamma} \dv_\beta v_\gamma$ is 
the vorticity field. We obtain, using (\ref{tpc}), the correlator
\bea
&&\langle  f_\alpha^\star (\vec x,t)  f_\beta^\star (\vec x',t') \rangle
\equiv  D_{\alpha \beta}(|\vec x -\vec x'|) \delta(t-t') \nonumber \\
&=& \epsilon_{\alpha \rho \sigma} \epsilon_{\beta \gamma \eta}
\dv_\rho \dv_\gamma' \langle f_\sigma (\vec x,t) f_\eta (\vec x',t')\rangle \nonumber \\
&=&(\dv_\alpha \dv_\beta - \delta_{\alpha \beta} \dv^2)
F(|\vec x - \vec x'|) \delta(t-t') \ . \ \label{tpcb}
\eea

We state now, in field theoretical language, what is meant by the stochastic evolution problem. Defining $\omega_\alpha^0 (\vec x)$ as the vorticity field at a certain time instant $t_0$, we are interested to find the probability density functional $Z=Z[\omega_\alpha (\vec x),t_1|\omega_\alpha^0 (\vec x),t_0]$ for the observation of vorticity $\omega_\alpha (\vec x)$ at a latter time instant $t_1$. Within the path-integral version of the Martin-Siggia-Rose formalism \cite{msr}, it follows, from Eqs. (\ref{sh-eq}) and (\ref{tpcb}), that
\be
Z= {\cal{N}}\int D \hat \omega_\alpha D \omega_\alpha \exp( i S) \ , \ \label{msr-pdf}
\ee
where ${\cal{N}}$ is a normalization constant \cite{comment}, to assure that
$\int D  \omega_\alpha (\vec x) Z[\omega_\alpha (\vec x),t_1|\omega_\alpha^0 (\vec x),t_0] =1$,
and
\bea
S &=& \int_{t_0}^{t_1} dt \{ \int d^3 \vec x \hat \omega_\alpha (\dv_t \omega_\alpha
+ v_\beta \dv_\beta \omega_\alpha - \omega_\beta \dv_\beta v_\alpha
- \nu \dv^2 \omega_\alpha) \nonumber \\
&+& i \int d^3 \vec x d^3 \vec x'
\hat \omega_\alpha (\vec x,t) D_{\alpha \beta}(|\vec x -\vec x'|)
\hat \omega_\beta (\vec x',t) \} \ . \ \label{msr-action}
\eea
An effective model of turbulent dynamics would be naturally attained if the velocity and vorticity fields that appear in (\ref{msr-action}) were expressed as a sum over the contributions produced exclusively by relevant flow profiles. The basic difficulty here regards the selection and parametrization of such configurations. A promising starting point is to establish a set of ``building blocks" that could be used to represent the usually observed coherent structures. We recall that vortex sheets or tubes, in particular, can be exactly obtained in a simple way as linear combinations of elementary closed vorticity rings, through a lattice vortex construction \cite{tagu, chorin} originally devised in the realm of superfluid physics \cite{williams}. Just define a cubic lattice, with spacing parameter $\epsilon \rightarrow 0$ (i.e., much smaller than the Kolmogorov dissipation length $\eta$), whose sites are written as $\vec x_p = \epsilon (p_1 \hat x_1 +p_2 \hat x_2 + p_3 \hat x_3)$, where the $p_i$'s are integers. The vector position $\vec x_p$ is taken to be the common vertex of three plaquettes oriented according to the unit vectors $\hat x_\sigma$, as shown in Fig. 1. In a self-evident notation, an arbitrary plaquette is completely charaterized by the vector doublet ${\cal{P}} = (\vec x_p,\hat x_\sigma)$. Furthermore, by definition, the plaquette's boundary $\partial {\cal{P}}$ is identified to a line vortex (vortex tube with vanishing cross section) which carries vorticity flux $\phi_\sigma (\vec x_p,t)$. Of course, a square line vortex is an ill-defined mathematical object, due to the divergence of the velocity field on its corners \cite{saffman}. However, as a simple regularization procedure, we impose an ultraviolet cutoff $ \Lambda  \equiv 1 / \epsilon$ in the Fourier-transformed kernel of the operator that maps vorticity into velocity. In rough terms, this is equivalent to replace the line vortex by a vortex tube with a cross section of radius $\sim \epsilon$. It is important to remark that the lattice vortex is in fact an ``overcomplete" basis for the description of general flow configurations. 

Considering the plaquette ${\cal{P}} = (\vec x_p,\hat x_\sigma)$, let $x_\alpha(s) \in \dv {\cal{P}}$ be a point parametrized by the arclength $0 \leq s \leq 4 \epsilon$ of the oriented boundary path that starts at the reference point $\vec x_p$ and ends at $x_\alpha(s)$. The vorticity field associated with this plaquette is
\be
\omega_\alpha  = \phi_\sigma (\vec x_p,t)
\delta (n_1) \delta (n_2) { d \over {d s}} x_\alpha (s) \ , \ \label{loop}
\ee
where $n_1$ and $n_2$ are the coordinates along the normal and binormal directions on the line vortex (the binormal vector is defined as $\hat x_\sigma$). The central idea underlying the lattice vortex representation is, then, to substitute (\ref{loop}) in the Martin-Siggia-Rose action (\ref{msr-action}) and perform afterwards the sum over all the plaquettes. Introducing $F_{\alpha \beta} \equiv \dv_\alpha \hat \omega_\beta - \dv_\beta \hat \omega_\alpha$, we get, for a single plaquette, the following relations:
\bea
&&\int d^3 \vec x \hat \omega_\alpha \dv_t \omega_\alpha
= \dv_t \phi_\sigma (\vec x_p,t) \oint_{\dv \cal{P}} dx_\alpha \hat \omega_\alpha = - \dv_t \phi_\sigma (\vec x_p,t) \oint_{\dv \cal{P}} dx_\alpha
\dv^{-2} \dv_\beta F_{\alpha \beta} \ , \ \nonumber \\
&&\int d^3 \vec x \hat \omega_\alpha (v_\beta \dv_\beta \omega_\alpha
- \omega_\beta \dv_\beta v_\alpha ) = \phi_\sigma (\vec x_p,t) \oint_{\dv \cal{P}} dx_\alpha F_{\alpha \beta} v_\beta \ , \ \nonumber \\
&&\int d^3 \vec x \hat \omega_\alpha \dv^2 \omega_\alpha =
\phi_\sigma (\vec x_p,t) \oint_{\dv \cal{P}} dx_\alpha \dv^2 \hat \omega_\alpha = - \phi_\sigma (\vec x_p,t) \oint_{\dv \cal{P}} dx_\alpha \dv_\beta F_{\alpha \beta} \ . \
\eea
Thus, the Martin-Siggia-Rose action becomes
\bea
S &=&  \int_{t_0}^{t_1} dt \{ \sum_{\cal{P}} \oint_{\dv \cal{P}} dx_\alpha
[ \dv_t \phi_\sigma \dv^{-2} \dv_\beta F_{\alpha \beta} +  \phi_\sigma F_{\alpha \beta} v_\beta
+ \nu \phi_\sigma \dv_\beta F_{\alpha \beta} ] \nonumber \\
&+& { i \over 2} \int d^3 \vec x d^3 \vec x'
F_{\alpha \beta} (\vec x,t) F(| \vec x - \vec x'|) F_{\alpha \beta}(\vec x',t) \} \ . \ \label{msr-actionb}
\eea
It is worth noting that the action (\ref{msr-actionb}) is invariant under the local transformation $\hat \omega_\alpha \rightarrow \hat \omega_\alpha + \dv_\alpha \chi$. In fact, it turns out that the transition probability for small time intervals, derived from (\ref{msr-pdf}), with (\ref{msr-actionb}), is well approximated by the expectation value of a product of loop operators, computed in a non-local, three-dimensional, $U(1)$ gauge theory. 

Drawning upon the gauge field theory correspondence, we define now the dual field strength,
\be
\hat \phi_\alpha (\vec x,t) \equiv {1 \over 2} \epsilon_{\alpha \beta \gamma} F_{\beta \gamma}
(\vec x,t) \ , \
\ee
which satisfies to $\dv_\alpha \hat \phi_\alpha =0$ and, additionally,
\bea
&&F_{\alpha \beta} = \epsilon_{\alpha \beta \gamma} \hat \phi_\gamma \ , \ \nonumber \\
&&\frac{1}{2} F_{\alpha \beta} F_{\alpha \beta} = \hat \phi_\alpha \hat \phi_\alpha
\ , \ \nonumber \\
&&dx_\alpha F_{\alpha \beta} v_\beta
= \epsilon_{\alpha \beta \gamma}  dx_\alpha  v_\beta \hat \phi_\gamma \ . \
\eea
We find, substituting the above relations in (\ref{msr-actionb}),
\bea
S &=& - \int_{t_0}^{t_1} dt  \{ \sum_{\cal{P}} \oint_{\dv \cal{P}} dx_\alpha \epsilon_{\alpha \beta \gamma}
[  \dv_t \phi_\sigma (\vec x_p,t) \dv^{-2} \dv_\beta \hat \phi_\gamma +  \phi_\sigma (\vec x_p,t) \hat \phi_\beta v_\gamma \nonumber \\
&-& \nu \phi_\sigma (\vec x_p,t) \dv_\beta \hat \phi_\gamma ]
+ i  \int  d^3 \vec x d^3 \vec x' \hat \phi_\alpha (\vec x,t) F(| \vec x - \vec x'|)
\hat \phi_\alpha (\vec x',t) \} \ .  \
\eea
The solenoidal constraint for the dual field can be imposed in the path-integration (\ref{msr-pdf}) by means of an auxiliary scalar field $\lambda$, which is nothing but a Lagrange multiplier. More concretely, we take
\be
Z= {\cal{N}}\int D \hat \phi_\alpha D \phi_\alpha D \lambda \exp( i S) \ , \ \label{msr-pdfb}
\ee
with
\bea
S &=& \int_{t_0}^{t_1} dt \{ \int d^3 \vec x (L_\alpha ( \vec x,t) + \dv_\alpha \lambda) \hat \phi_\beta (\vec x,t) \nonumber \\
&+& i \int d^3 \vec x d^3 \vec x'
\hat \phi_\alpha (\vec x,t) F(| \vec x - \vec x'|)
\hat \phi_\alpha (\vec x',t)  \} \ .  \ \label{msr-actionc}
\eea
In (\ref{msr-actionc}), the whole dependence on the $\phi$-fields is implicit in the non-local ``source" term
\bea
L_\alpha (\vec x,t) &\equiv& - \sum_{\cal{P}} \oint_{\dv \cal{P}} dx_\gamma'
\epsilon_{\alpha \beta \gamma}[ \dv_t \phi_\sigma (\vec x_p,t) \dv^{-2} \dv_\beta \nonumber \\
&+&  \phi_\sigma (\vec x_p,t)v_\beta (\vec x',t)
- \nu \phi_\sigma (\vec x_p,t) \dv_\beta  ] \delta^3(\vec x'-\vec x) \ .\
\eea
To proceed, we define the Fourier transform of $L_\alpha (\vec x,t)$,
\bea
&&\tilde L_\alpha (\vec k,t) = \int d^3 \vec x \exp(-i \vec k \cdot \vec x) 
L_\alpha (\vec x,t) =  i \sum_{\cal{P}} \oint_{\dv \cal{P}} dx_\gamma' \epsilon_{\alpha \beta \gamma}
\exp(-i \vec k \cdot \vec x') \nonumber \\
&&\times [ { k_\beta \over k^2} \dv_t \phi_\sigma (\vec x_p,t)
+ i \phi_\sigma (\vec x_p,t) v_\beta (\vec x',t) + \nu \phi_\sigma (\vec x_p,t) k_\beta] \ . \ 
\eea
Writing, for a given plaquette ${\cal{P}} = (\vec x_p,\hat x_\sigma)$, the boundary position vector as $\vec x' = \vec x_p + \vec \xi$, with $v_\alpha (\vec x') \simeq v_\alpha (\vec x_p)+\xi_\eta \dv_\eta v_\alpha (\vec x_p)$, we obtain
\bea
&&\tilde L_\alpha (\vec k,t) = 
i \sum_{p} \exp(-i \vec k \cdot \vec x_p) \epsilon_{\alpha \beta \gamma} 
\{ g_{\sigma \gamma} (\vec k) [ {  k_\beta \over k^2} \dv_t \phi_\sigma (\vec x_p,t) \nonumber \\
&&+ i \phi_\sigma (\vec x_p,t) 
v_\beta (\vec x_p,t) + \nu  k_\beta \phi_\sigma (\vec x_p,t) ]+ g_{\sigma \gamma , \eta} (\vec k) \phi_\sigma (\vec x_p,t) \dv_\eta v_\beta (\vec x_p,t) \} \ , \
\eea
with 
\bea
&&g_{\sigma \gamma} (\vec k) \equiv  \oint_{\dv \cal{P}} d \xi_\gamma
\exp(-i \vec k \cdot \vec \xi) \ , \ \nonumber \\
&&g_{\sigma \gamma , \eta} (\vec k) \equiv  -i \oint_{\dv \cal{P}} d \xi_\gamma \xi_\eta
\exp(-i \vec k \cdot \vec \xi) = \frac{\dv}{\dv k_\eta} g_{\sigma \gamma} (\vec k) \ . \
\eea
In the limit $\epsilon \rightarrow 0$, keeping $k \ll 1 / \epsilon$, we have, asymptotically, $g_{\alpha \beta} (\vec k) = i \epsilon^2 
\epsilon_{\alpha \beta \gamma} k_\gamma$, 
and, therefore,
\bea
&&\tilde L_\alpha (\vec k,t) =
\epsilon^2 \sum_p \exp(-i \vec k \cdot \vec x_p) \{
\tilde \Pi_{\alpha \beta}  (\vec k) [ \dv_t \phi_\beta (\vec x_p,t) + \nu k^2 \phi_\beta (\vec x_p,t)] \nonumber \\
&&+ i \phi_\beta (\vec x_p,t) [ \delta_{\alpha \beta} v_\gamma (\vec x_p,t) 
k_\gamma  -i \dv_\alpha v_\beta (\vec x_p,t)- k_\alpha v_\beta (\vec x_p,t)] \}
\ , \ \label{fl}
\eea
where we used $\tilde \Pi_{\alpha \beta} (\vec k) = \delta_{\alpha \beta} - k_\alpha k_\beta / k^2$, the 
Fourier-transformed projector on transverse modes. The continuum limit of the above sum is defined through the substitutions
\be
\vec x_p \rightarrow \vec x \ , \
\sum_{p} \rightarrow \frac{1}{\epsilon^3} \int d^3 \vec x \ , \ 
\phi_\beta \rightarrow \frac{\phi_\beta}{\epsilon} \ . \ \label{cl}
\ee
We find
\be
\tilde L_\alpha (\vec k,t) = \tilde \Pi_{\alpha \beta} (\vec k) [ \dv_t \tilde \phi_\beta
(\vec k,t) + \nu k^2 \tilde \phi_\beta (\vec k,t)] + \Omega_\alpha (\vec k,t) \ , \
\ee
where 
\bea
\tilde \phi_\alpha (\vec k,t) &=& \int d^3 \vec x \exp(-i \vec k \cdot \vec x) \phi_\alpha (\vec x,t) 
\ , \ \nonumber \\
\Omega_\alpha (\vec k,t) &=& \int d^3 \vec x \exp(-i \vec k \cdot \vec x) v_\beta (\dv_\beta \phi_\alpha
-\dv_\alpha \phi_\beta ) \ . \
\eea
There is a simple connection between $\phi_\alpha$ and the velocity field $v_\alpha$. Taking the Fourier transform of the vorticity field, we find
\bea
&&\tilde \omega_\alpha (\vec k,t) =
\int d^3 \vec x \exp(-i \vec k \cdot \vec x) \omega_\alpha (\vec x)
\nonumber \\
&=& \sum_{\cal{P}} \oint_{\dv \cal{P}} dx_\alpha 
\exp(-i \vec k \cdot \vec x) \phi_\sigma (\vec x_p,t) \nonumber \\
&=& \sum_{p} g_{\beta \alpha} (\vec k)
\exp(-i \vec k \cdot \vec x_p) \phi_\beta (\vec x_p,t) \nonumber \\
&=& i \epsilon^2 \epsilon_{\alpha \beta \gamma} k_\beta  \sum_p
\exp(-i \vec k \cdot \vec x_p) \phi_\gamma (\vec x_p,t) \ . \
\eea
Recalling (\ref{cl}), we get $\omega_\alpha = \epsilon_{\alpha \beta \gamma} \dv_\beta \phi_\gamma$ in the continuum limit. Since $\omega_\alpha = \epsilon_{\alpha \beta \gamma} \dv_\beta v_\gamma$, we immediately conclude that the fields $v_\alpha$ and $\phi_\alpha$ differ only by a gradient, which means that $v_\alpha = \Pi_{\alpha \beta} \phi_\beta$. As a consequence, if (\ref{fl}) is taken back to real space, we get
\be
L_\alpha (\vec x,t) = \dv_t  v_\alpha - \nu \dv^2 v_\alpha + v_\beta (\dv_\beta \phi_\alpha
-\dv_\alpha \phi_\beta ) \ . \
\ee
We define at this point the additional scalar field $\zeta= [\dv_t  - \nu \dv^2 ]^{-1} \lambda$, and impose $\phi_\alpha = v_\alpha + \dv_\alpha \zeta$, so that
\be
L_\alpha + \dv_\alpha \lambda = \dv_t  \phi_\alpha - \nu \dv^2 \phi_\alpha + v_\beta (\dv_\beta \phi_\alpha
-\dv_\alpha \phi_\beta ) \ . \ \label{l-phi}
\ee
Since the action (\ref{msr-actionc}) is quadratic in $\hat \phi_\alpha$, it is possible to evaluate the exact path-integration over the dual fields. Using (\ref{msr-pdfb}), (\ref{msr-actionc}) and (\ref{l-phi}), the result is an effective (and exact) expression for the probability density functional $Z$,
\be
Z = {\cal{N}} \int D\phi_\alpha D \lambda \exp(i S_\phi) \ , \ \label{z-phi}
\ee
where
\bea
S_\phi &=&  \frac{i}{4} \int_{t_0}^{t_1} dt \int d^3 \vec x d^3 \vec x'  [\dv_t  \phi_\alpha - \nu \dv^2 \phi_\alpha 
+ v_\beta (\dv_\beta \phi_\alpha
- \dv_\alpha \phi_\beta )]_x  \nonumber \\
&\times& F^{-1} (\vec x - \vec x') [\dv_t  \phi_\alpha 
- \nu \dv^2 \phi_\alpha + v_\beta (\dv_\beta \phi_\alpha
-\dv_\alpha \phi_\beta )]_{x'} \ . \ \label{s-phi}
\eea
The field theory given by (\ref{s-phi}) may be obtained directly, along the Martin-Siggia-Rose formalism, from the stochastic differential equation
\be
\dv_t \phi_\alpha + v_\beta (\dv_\beta \phi_\alpha - \dv_\alpha \phi_\beta) =
\nu \dv^2 \phi_\alpha + f_\alpha \ . \ \label{sns-eqsb}
\ee
For vanishing external forces, the above expression reduces to the usual dipole version of the Navier-Stokes equation \cite{chorin}. Substituting $\phi_\alpha$ by $v_\alpha + \dv_\alpha \zeta$ in (\ref{sns-eqsb}), we get the original stochastic Navier-Stokes Eqs. (\ref{sns-eqs}), with pressure given by $P= \dv_t \zeta -\nu \dv^2 \zeta - \frac{1}{2} \vec v^2$.

Does Eq. (\ref{sns-eqsb}) yield any advantage over the standard Navier-Stokes formulation? Direct numerical simulations based on (\ref{sns-eqsb}) would probably have the same computational cost than the ones which usually rely on the Navier-Stokes equations, since both versions involve at least two Fourier transformations per iteration cycle. In practice, the above description provides an alternative approach to large eddy simulations \cite{tagu}, or the analysis of phenomenological aspects of vortex tube dynamics, as put forward in the following considerations. 

\section{Stochastic Vortex Tube Evolution}

We are now interested to investigate the evolution of a closed vortex tube $\Gamma$, with small linear cross-sectional dimensions (of the order of $\eta$) and subject to the action of large scale gaussian random forces. In a first approximation, we regard the tube as a vorticity filament, parametrized by the curve $x_\alpha = x_\alpha (s,t)$, and carrying total vorticity flux $\phi$. The vorticity field is given by
\be
\omega_\alpha  = \phi \delta (n_1) \delta (n_2) { d \over {d s}} x_\alpha (s,t) \ , \ \label{vtube}
\ee
where, similarly to the former plaquette's definitions, $n_1$ and $n_2$ indicate the normal and binormal coordinates along the line vortex. 

The assumptions taken in (\ref{vtube}) that the vorticity flux is time-independent and that cross-section fluctuations may be neglected are imposed as phenomenological constraints. Our results will be expected to hold to the extent that phenomena like vortex breakdown, vortex merging, etc. do not affect the vortex tube evolution. Such flow regimes have been well verified in the numerical and real experiments where vortices are mostly advected by the background flow, during their mean life-time, in agreement with the flux conservation Kelvin theorem. This state of affairs gives in fact the physical basis that supports the somewhat popular choice of modeling vortex tubes by means of Burgers vortices, or similar configurations.

Our first task here is to apply the information provided by (\ref{vtube}) in the effective action (\ref{s-phi}). In the limit of vanishing viscosity, we are left, therefore, with the evaluation of $\dv_t \phi_\alpha$ and $v_\beta (\dv_\beta \phi_\alpha -\dv_\alpha \phi_\beta )$. The latter quantity is just minus the Lamb vector. In fact, using $\omega_\alpha = \epsilon_{\alpha \beta \gamma} \partial_\beta \phi_\gamma$ a straigthforward computation
leads to 
\be
v_\beta (\dv_\beta \phi_\alpha -\dv_\alpha \phi_\beta ) = \epsilon_{\alpha \beta \gamma} \omega_\beta v_\gamma \ . \ \label{nlin}
\ee 
On the other hand, to find $\dv_t \phi_\alpha$, let us imagine as an auxiliary construction that the line vortex is advected by a divergence-free field $\xi_\alpha(\vec x,t)$, defined on all space, and which satisfies the boundary condition
$\xi_\alpha(\vec x ,t)= \dot x_\alpha(s,t)$ on $\Gamma$. We have, then, \cite{aches}
\be
\dv_t \vec \omega 
=\vec \nabla \times (\vec \xi \times \vec \omega) \label{lamb-eq}
\ . \
\ee
Observing that $\dv_t \omega_\alpha = 
\dv_t [\epsilon_{\alpha \beta \gamma} \dv_\beta \phi_\gamma] = 
\epsilon_{\alpha \beta \gamma} \dv_\beta \dv_t \phi_\gamma$, we get, from (\ref{lamb-eq}), 
\be
\partial_t \phi_\alpha = \epsilon_{\alpha \beta \gamma} \dot x_\beta \omega_\gamma + \partial_\alpha \lambda \ , \ \label{phi-d}
\ee
where $\lambda$ is an arbitrary field. We are ready to substitute (\ref{nlin}) and (\ref{phi-d}) into (\ref{s-phi}). Introducing
\be
\psi^{\perp}_\alpha(s,t) \equiv \epsilon_{\alpha \beta \gamma} \psi_\beta(s,t) 
\frac{d}{ds} x_\gamma(s,t)\ , \
\ee
where $\psi_\alpha=\dot x_\alpha - v_\alpha$, we obtain
\be
S_\phi = S_{\psi \psi}+S_{\lambda \psi}+S_{\lambda \lambda} \ , \
\ee
with
\bea
S_{\psi \psi} &=& \frac{i \phi^2}{4} \int_{t_0}^{t_1} dt \int_0^{p(t)} ds \int_0^{p(t)} ds' \psi^{\perp}_\alpha(s,t) F^{-1}(\vec x(s) -\vec x(s')) \psi^{\perp}_\alpha (s',t) \ , \ \nonumber \\
S_{\lambda \psi} &=& \frac{i \phi}{2} \int_{t_0}^{t_1} dt \int d^3 \vec x \int_0^{p(t)} ds' \dv_\alpha \lambda (\vec x,t) F^{-1}(\vec x -\vec x(s'))  \psi^{\perp}_\alpha (s',t) \ , \ \nonumber \\
S_{\lambda \lambda} &=& \frac{i}{4} \int_{t_0}^{t_1} dt \int d^3 \vec x \int d^3 \vec x' \dv_\alpha \lambda (\vec x,t) F^{-1}(\vec x -\vec x')  \dv_\alpha \lambda (\vec x',t) \ , \
\eea
where $p(t)$ is the length of the vorticity filament. The integration over $\lambda$ gives
\be
Z = {\cal{N}} \int D \psi_\alpha^\perp \exp( i S_\psi ) \ , \
\ee
where
\be
S_\psi = \frac{i \phi^2}{4} \int_{t_0}^{t_1} dt \int_0^{p(t)} ds \int_0^{p(t)} ds' \psi^{\perp}_\alpha(s,t) 
\Pi_{\alpha \beta} F^{-1}(\vec x(s) -\vec x(s')) \psi^{\perp}_\alpha (s',t) \ . \ \label{s-psi}
\ee
Note that while $S_\psi$ is a functional of $\psi_\alpha^\perp$, the projection of $\psi_\alpha$ on $d x_\alpha / ds$ (that is, the longitudinal component of $\psi_\alpha$) maps the line vortex into itself. The singularities that eventually appear in integrand of (\ref{s-psi}) may be circumvented in a physical way, replacing the original vortex filament by a vortex tube, through the substitutions
\bea
&&\psi_\alpha^\perp (s,t) \rightarrow \psi_\alpha^\perp (s,t) h(n_1,n_2) \ , \ \nonumber \\
&&\psi_\alpha^\perp (s',t) \rightarrow \psi_\alpha^\perp (s',t) h(n_1',n_2') \ , \ \nonumber \\
&&ds \rightarrow d^3 \vec x \ , \ ds' \rightarrow d^3 \vec x'  \ , \ \nonumber \\
&&F^{-1}(\vec x(s) -\vec x(s')) \rightarrow F^{-1}(\vec x -\vec x') \ , \ \label{reg}
\eea
where $\vec x = (n_1,n_2,s)$, $\vec x' = (n_1',n_2',s')$, and
\be
h(n_1,n_2) = \frac{1}{\pi \eta^2} \exp[ -\frac{1}{\eta^2}(n_1^2+n_2^2)] \ . \ \label{regb}
\ee
We assume that the curvature radius of the vortex tube is much larger than the Kolmogorov dissipation length, an hypothesis supported by observations. A necessary condition for this property to be preserved in time is
\be 
\eta | \dv_s \vec \psi^\perp| \ll |\vec \psi^\perp| \ . \ \label{smooth-vt}
\ee
In practical computations, we may work with a straight vortex tube, taking $s=z$, $n_1=x$, and $n_2=y$ (one can figure it out as a circular vortex tube with infinite curvature radius). We deal below with two specific examples of external stochastic forcing, given by (\ref{<ff>}) and (\ref{<ffb>}), which will be named models A and B, respectively. A more concise expression for $S_\psi$, compared to (\ref{s-psi}), follows in general, relying basically on
the slender profile of the vortex tube.
\vspace{0.2cm}

{\leftline{{\it{Analysis of Model A.}}}}
\vspace{0.2cm}

To evaluate $S_\psi$ it is necessary to consider the operator kernel,
\be
\Pi_{\alpha \beta} F^{-1}(\vec x -\vec x')=
\frac{1}{8 \pi D_0 m} (\delta_{\alpha \beta} - \dv^{-2} \dv_\alpha \dv_\beta)
(\partial^2 -m^2)^2 \delta^3(\vec x - \vec x') \ . \ \label{tilde-pi-f}
\ee
If this expression is substituted into (\ref{s-psi}), considering (\ref{reg}) and (\ref{regb}), a number of terms is obtained, hierarchically organized according to the powers of the dissipation length $\eta \rightarrow 0$ defined in their coeficients. We will retain in the expression for $S_\psi$ only the dominant term, corresponding to the smallest power of $\eta$. Using rotation invariance around the $z$ axis, and neglecting derivatives of $\psi^\perp_\alpha$ along the $z$ direction, as it follows from (\ref{smooth-vt}), this prescription effectively amounts to perform in (\ref{s-psi}) the replacement
\be
\Pi_{\alpha \beta} F^{-1}(\vec x -\vec x') \rightarrow \frac{\delta_{\alpha \beta}}{16 \pi D_0 m} (\dv_\perp^2)^2 \delta^3(\vec x - \vec x')
\ , \
\ee
where $(\dv_\perp)^2 \equiv \dv_x^2+\dv_y^2$. We obtain
\be
S_\psi = \frac{i \phi^2}{16 \pi^2 D_0 m \eta^6} \int_{t_0}^{t_1} dt \int_0^{p(t)} ds [ \psi^\perp_\alpha (s,t)]^2 \ . \  \label{s-psib}
\ee
Although (\ref{s-psib}) is an apparently simple quadratic action, the time-dependent spatial integration limit $p(t)$ renders the analytical evaluation of $Z$ difficult. Nevertheless, the problem looks amenable of numerical investigation through the use of Langevin techniques \cite{mori-nob}.

Since we are discussing the time evolution of a vortex tube, the relevant physical question one may ask is concerned with the probability density functional of finding the tube in a certain geometrical configuration. In a first instance, this seems to be an intricate problem, once any individual vortex tube ``world-line" to be considered in the path integration is accounted for by a large number of configurations of $\psi_\alpha$. A simple solution of this degeneracy problem may be obtained, however, by means of the ``minimal mapping" $\psi^0_\alpha$, depicted in Fig. 2. The essential idea is to keep track of the vortex tube evolution for a very small time interval $\delta$. We decompose the time evolution in two steps. First, the tube $\Gamma(t)$ is mapped into $\Gamma^\ast (t)$ through its self-induced velocity field $v_\alpha$. Next, the stochastic perturbation $\psi_\alpha$ takes $\Gamma^\ast (t)$ to the final configuration $\Gamma(t+\delta)$. The mapping sequence is $x_\alpha \rightarrow  x_\alpha' \rightarrow  x_\alpha''$, with
\bea
x_\alpha' &=& x_\alpha + \delta v_\alpha \ , \  \nonumber \\
x_\alpha''&=& x_\alpha' + \delta \psi_\alpha \ . \
\eea
Let $\gamma$ be the plane that contains $x_\alpha'$ and is normal to $\Gamma^\ast (t)$. Then, $\psi^0_\alpha$ is just the vector parallel to $\gamma$ that connects $x_\alpha'$ to the vortex tube $\Gamma(t+\delta)$. We have
\bea
&&\psi_\alpha^\perp(s,t) = \psi_\alpha^0(s+\delta \psi_s(s,t),t) + O(\delta^2)\nonumber \\
&&= \psi_\alpha^0(s,t) + \delta \psi_s \dv_s \psi_\alpha^0 (s,t) + O(\delta^2) \ , \ \label{exps-psi}
\eea
where $\psi_s \equiv  \psi_\alpha d x_\alpha  / ds$. The expansion (\ref{exps-psi}) implies that $\psi^\perp_\alpha-\psi^0_\alpha = O(\delta)$, and so $\psi^\perp_\alpha$ may be substituted by $\psi^0_\alpha$ in (\ref{s-psib}). We find, thus, that the probability density functional for the transition $\Gamma(t_0) \rightarrow \Gamma(t_1)$ of the vortex tube configuration may be defined as
\be
Z_A = {\cal{N}}\int D \psi^0_\alpha \exp \{ - \frac{\phi^2}{16 \pi^2 D_0 m \eta^6}  \int_{t_0}^{t_1} dt \int_0^{p(t)} ds [ \psi^0_\alpha(s,t)]^2 \} \ . \ \label{pdf-A}
\ee
Clearly, the original field degeneracy is removed, and there is in (\ref{pdf-A}) an one-to-one correspondence between the vortex tube integration paths and the fields $\psi^0_\alpha (s,t)$.
\vspace{0.2cm}

{\leftline{{\it{Analysis of Model B.}}}}
\vspace{0.2cm}

The computational steps are exactly the same as the ones performed in the former case. The only technical difference is that the analogous of 
Eq. (\ref{tilde-pi-f}) is written now in Fourier space:
\be
\tilde \Pi_{\alpha \beta} \tilde F^{-1}(k)=
\frac{1}{2 \pi^2 D_0}( \delta_{\alpha \beta} k^3-k_\alpha k_\beta k) \ . \
\ee
The dominant contribution to (\ref{s-psi}), of order $1 / \eta^5$, comes from the substitution 
\be
\tilde \Pi_{\alpha \beta} \tilde F^{-1}(k) \rightarrow
\frac{1}{4 \pi^2 D_0} \delta_{\alpha \beta} k_\perp^3 \ . \
\ee
We get, similarly to (\ref{pdf-A}), the probability density functional
\be
Z_B = {\cal{N}}\int D \psi^0_\alpha \exp \{ - \frac{6 \phi^2 \sqrt{\pi}}{D_0 \eta^5} \int_{t_0}^{t_1} dt \int_0^{p(t)} ds [ \psi^0_\alpha(s,t)]^2 \} \ . \ \label{pdf-B}
\ee
A remarkable feature of model B, as it may be easily inferred from (\ref{pdf-B}), is that there is no dependence of the probability density functional $Z_B$ upon the integral scale $L=1/m$ (as it occurs in model A, for instance).

In order to establish a connection between the above models and observed features of turbulent flows, a slight modification of expressions (\ref{pdf-A}) and (\ref{pdf-B}) is necessary. In principle, the Martin-Siggia-Rose framework implemented by (\ref{z-phi}) and (\ref{s-phi}) is expected to provide a bona fide statistical modeling of vortex tube motion if an ultraviolet cutoff appears dynamically at a frequency $|\omega | \sim 1/t_\eta$, where $t_\eta \sim \eta^{2/3}$ is the eddy turnover time at the Kolmogorov length scale. The simplest way to find improved versions of (\ref{pdf-A}) and (\ref{pdf-B}), thus, is to replace the Dirac's delta factor in (\ref{tpc}) by a regularized expression like
\be
\delta_R (t-t') = \frac{1}{2 t_\eta} \exp(-t_\eta^{-1}|t-t'|) \ , \ \label{reg-delta}
\ee
and relax the cutoff prescription for the field $\psi_\alpha$ in frequency space. As a consequence, if all the steps leading to (\ref{pdf-A}) and (\ref{pdf-B}) are evaluated again, taking into account the modifications due to (\ref{reg-delta}), we will get, for both models A and B, the general result
\be
Z = {\cal{N}}\int D \psi^0_\alpha \exp \{ - c \int_{t_0}^{t_1} dt \int_0^{p(t)} ds
[ (t_\eta \dv_t \psi^0_\alpha)^2 + 
(\psi^0_\alpha)^2 ] \} \ , \  \label{g-z}
\ee
where the constant $c=c(m,\eta,D_0)$ is defined in the same way as before. The spatial wavenumber cutoff is hidden in (\ref{g-z}), insofar as it will not have any relevant role in the forthcoming arguments [the cutoff is much smaller than $k_\eta \sim 1 / \eta$, according to (\ref{smooth-vt})].

\section{Background Velocity Fluctuations}

It is interesting to note that the probability density functional (\ref{g-z}) is completely equivalent to the one derived for the problem
of random advection of a vortex tube by a background velocity field. In this way, we can draw a correspondence between the former effective
description, based on the analysis of the stochastic Navier-Stokes equations, and realistic properties of turbulent flows. In more precise words, let $v_\alpha(\vec x,t)$ be the velocity of the background flow, which is assumed to be a random gaussian fluctuating field, with vanishing mean value and correlator
\be
\langle v_\alpha(\vec x,t) v_\beta (\vec x' ,t') \rangle = g \Pi_{\alpha \beta} \delta^3(\vec x - \vec x') \delta_R(t-t') \ . \ \label{vv}
\ee
It follows, therefore, that the one-dimensional background energy spectrum is given by
\be
E(k)= \frac{g k^2}{4 \pi^2 t_\eta} \ , \ \label{spectrum}
\ee
and that the path-integral expression (\ref{g-z}) holds, with 
\be
c= \frac{\pi \eta^2}{2 g} \ . \ \label{c-value}
\ee
It is straightforward to prove (\ref{spectrum}) from the Fourier transform of the velocity-velocity correlator (\ref{vv}). 
Below, we discuss in more detail how (\ref{g-z}) arises from (\ref{vv}), with the specific parameter definition (\ref{c-value}). 
Other velocity correlators would work as well. However, (\ref{vv}) is particularly attractive in view of its direct relation to
numerical observations \cite{farge}.

The probability density functional to have a certain background velocity field $\bar v_\alpha(\vec x,t)$ in the region $\Omega_t$ enclosed by a vortex tube, for the time interval $t_0 \leq t \leq t_1$, may be written as
\be
{\cal{P}}= \langle \Pi_{i,j} \delta( \bar v_\alpha(\vec x_i,t_j) - v_\alpha(\vec x_i,t_j)) \rangle \ ,\ \label{v-pdf}
\ee
where $(\vec x_i,t_j)$ denotes a discretized space-time position defined in the set of world-lines generated by the vortex tube evolution.
Using the Fourier representation of the delta function, Eq. (\ref{v-pdf}) becomes, in the continuum limit, 
\be
{\cal{P}} = {\cal{N}} \int D \xi_\alpha \exp( i \int_{t_0}^{t_1} dt \int_{\Omega_t} d^3 \vec x \xi_\alpha \bar v_\alpha)
 \langle \exp( -i \int_{t_0}^{t_1} dt \int_{\Omega_t} d^3 \vec x
\xi_\alpha v_\alpha) \rangle \ . \
\ee
Resorting to the gaussian random behavior of the background velocity field, we are able to compute the above expectation value. Using (\ref{vv}) we find,
\bea
{\cal{P}} &=& {\cal{N}} \int D \xi_\alpha \exp( i \int_{t_0}^{t_1} dt \int_{\Omega_t} d^3 \vec x \xi_\alpha \bar v_\alpha)
\exp[ -\frac{g}{2} \int_{t_0}^{t_1} dt \int_{t_0}^{t_1} dt' \nonumber \\
&\times& \int_{\Omega_t} d^3 \vec x
\xi_\alpha (\vec x,t) \delta_R(t-t')\Pi_{\alpha \beta} \xi_\beta (\vec x,t')]\ . \
\eea
Since $\bar v_\alpha = \Pi_{\alpha \beta} \bar v_\beta$, we may integrate over the field $\xi_\alpha$ to get
\be
{\cal{P}} \propto \exp \{ - \frac{1}{2g} \int_{t_0}^{t_1} dt \int_{\Omega_t} d^3 \vec x 
[ (t_\eta \dv_t \bar v_\alpha)^2 + 
(\bar v_\alpha)^2 ] \} \ . \ \label{v-pdfb}
\ee
If the vortex tube has a small circular cross section of area $\pi \eta^2$, we can replace $\int_{\Omega_t} d^3 \vec x$ by
$\pi \eta^2 \int_0^{p(t)} ds$ in (\ref{v-pdfb}). Furthermore, to find the transition probability density functional $Z$ for the vortex tube
evolution between configurations $\Gamma(t_0)$ and $\Gamma(t_1)$, we (i) decompose the velocity field in transverse and longitudinal components to the vortex tube tangent vector, viz., $\bar v_\alpha = \bar v^\perp_\alpha + \bar v^l_\alpha$, (ii) integrate over the longitudinal components $\bar v^l_\alpha$, and (iii) introduce the ``minimal velocity field" $v^0_\alpha$ in close analogy with the previous definition of $\psi^0_\alpha$. We obtain
\be
Z = {\cal{N}}\int D v^0_\alpha \exp \{ - \frac{\pi \eta^2}{2 g} \int_{t_0}^{t_1} dt \int_0^{p(t)} ds
[ (t_\eta \dv_t v^0_\alpha)^2 + 
(v^0_\alpha)^2 ] \} \ . \
\ee
Therefore, identifying $v_\alpha^0$ to $\psi^0_\alpha$, we have just found (\ref{g-z}) again, with $c$ given by (\ref{c-value}). From 
(\ref{spectrum}) and (\ref{c-value}), we can predict the form of the one-dimensional energy spectrum for models A and B (disregarding numerical prefactors):
\bea
E_A(k)&\sim& \frac{D_0 m \eta^8}{\phi^2 t_\eta} k^2 \ , \ \nonumber \\
E_B(k)&\sim& \frac{D_0 \eta^7}{\phi^2 t_\eta} k^2 \ . \
\eea
It is useful to compare the Kolmogorov's spectrum $E_K(k) \sim D_0^{2/3} k^{-5/3}$ with the above expressions. We may estimate, relying on Kolmogorov phenomenology, that $\phi \sim D_0^{1/3} \eta^{4/3}$ and $t_\eta \sim D_0^{-1/3} \eta^{2/3}$. At the dissipative wavenumber $k_\eta \sim 1/ \eta$, we define the Reynolds number dependent dimensionless ratio 
\be
Q \equiv \frac{E(k_\eta)}{E_K(k_\eta)} \sim R_e^\alpha \ , \ \label{q-factor}
\ee
where $E(k_\eta)$ is the background spectrum for a given model. It turns that for model A, we get $\alpha=-1$ while for model B, $\alpha=0$. More generally, it is not difficult to realize that the family of gaussian stochastic forces described by
\be
\tilde F(k) \sim (k^2 + m^2)^{-\beta} \ , \ \label{ffbeta}
\ee
with $\beta \geq 3/2$ leads to (\ref{q-factor}) with $\alpha = 3 - 2 \beta$.

A numerical wavelet analysis by Farge et al. \cite{farge} of the direct numerical simulations carried out by Vincent and Meneguzzi \cite{vinc} at moderately high Reynolds numbers, reveals the existence of a background $k^2$ one-dimensional energy spectrum. The turbulent flow may be depicted as a vortex tube gas surrounded by incoherent fluctuations, the latter having their kinetic energy equiprobably distributed over the spatial Fourier modes. It has been suggested in Ref. \cite{farge} that the dissipation at the bottom of the inertial range would be preceded at larger scales by some coherent-to-incoherent energy transfer from the vortex tubes to the background field. A fraction of the vortex tubes would be disrupted in a conservative way, so that the transformation of their mechanical energy into heat would occur afterwards from the background flow. One may conjecture that the integral length scale is irrelevant in this sequence of small scale events. In that case, we have $\alpha=0$, as in model B, which is actually the scenario indicated by the numerical results, where $Q \simeq 0.1$ for the Taylor-scale Reynolds number $R_\lambda = 150$ (equivalent to $R_e \simeq 10^4$, according to Lohse \cite{lohse}).
\vspace{-0.2cm}

\section{Conclusion}

We investigated in this work both formal and phenomenological aspects of homogenous isotropic turbulence, within the stochastic modeling of vorticity dynamics. A rigorous statistical lattice vortex description of turbulent flows was established, yielding the basis for a subsequent phenomenological discussion of the problem of the random evolution of vortex tubes, commonly observed in experiments and numerical simulations. Since the advection of vorticity coherent structures is ultimately caused by the background flow, according to Kelvin's theorem, we interpret the stochastic method as an effective tool for computing the evolution of vortex tubes. We were able to find in this way a plausible form for the background velocity-velocity correlator, and, as an immediate consequence, the background one-dimensional energy spectrum. We found a satisfactory agreement with the recent numerical analysis of Farge et al. \cite{farge}, where a thermal-like spectrum was clearly noticed for the background flow. In particular, we observed that the gaussian correlator (\ref{<ffb>}), used in the renormalization group approach to turbulence \cite{fns,d-m,y-o,smith} is likely the correct choice (model B of Sec. IV) for the derivation of phenomenologically meaningful results. Furthermore, it would be important to improve the connection between the stochastic modeling and the numerical results concerned with anisotropic effects, as the reported zero helicity distribution peak for the incoherent fluctuations \cite{farge}. 

There is a strong numerical evidence that the vortex tube gas accounts on its own for the Kolmogorov's spectrum \cite{hata,farge,araki,kivo}. Regarding the background flow, our analysis suggests that it has a twofold character, involving the combination of the ``eddy noise" \cite{frisch} forcing, effectively modeled by (\ref{<ffb>}), and of configurations which satisfy the energy equipartion principle. The picture that emerges -- to be explored in further analytical and numerical works -- is that these two facets of the background fluctuations are self-consistently related to the vorticity coherent structures. While the force-force correlator (\ref{ffbeta}), with $\beta > 3/2$ is a reasonable choice for a rigorous study of the turbulence problem, it becomes useless when considered in the simplified phenomenological perspective addressed in Sec. IV. On the other hand, model B is favored by the force of numerical observations, since it copes well with the tripartite phenomenological stage set up by the vortex tube gas, stochastic eddy noise, and the thermal-like background flow.


\newpage 

\begin{figure}
FIGURE CAPTIONS

\caption{
The three oriented plaquettes which have the common reference 
position $\vec x_p$, and carry, on their boundaries, vorticity 
fluxes $\Phi_1$, $\Phi_2$, and $\Phi_3$.}

\caption{
The vortex tube $\Gamma(t)$ evolves, during the small time interval 
$\delta$, to the new configuration $\Gamma(t+\delta)$. The intermediate dashed tube $\Gamma^\ast (t)$ corresponds to the transport provided by the self-induced velocity field $v_\alpha$.}

\end{figure}


\begin{thebibliography}{21}
\bibitem{comment} To save notation, we denote the other normalization constants defined throughout this work by the same symbol ${\cal{N}}$.
\end{thebibliography}
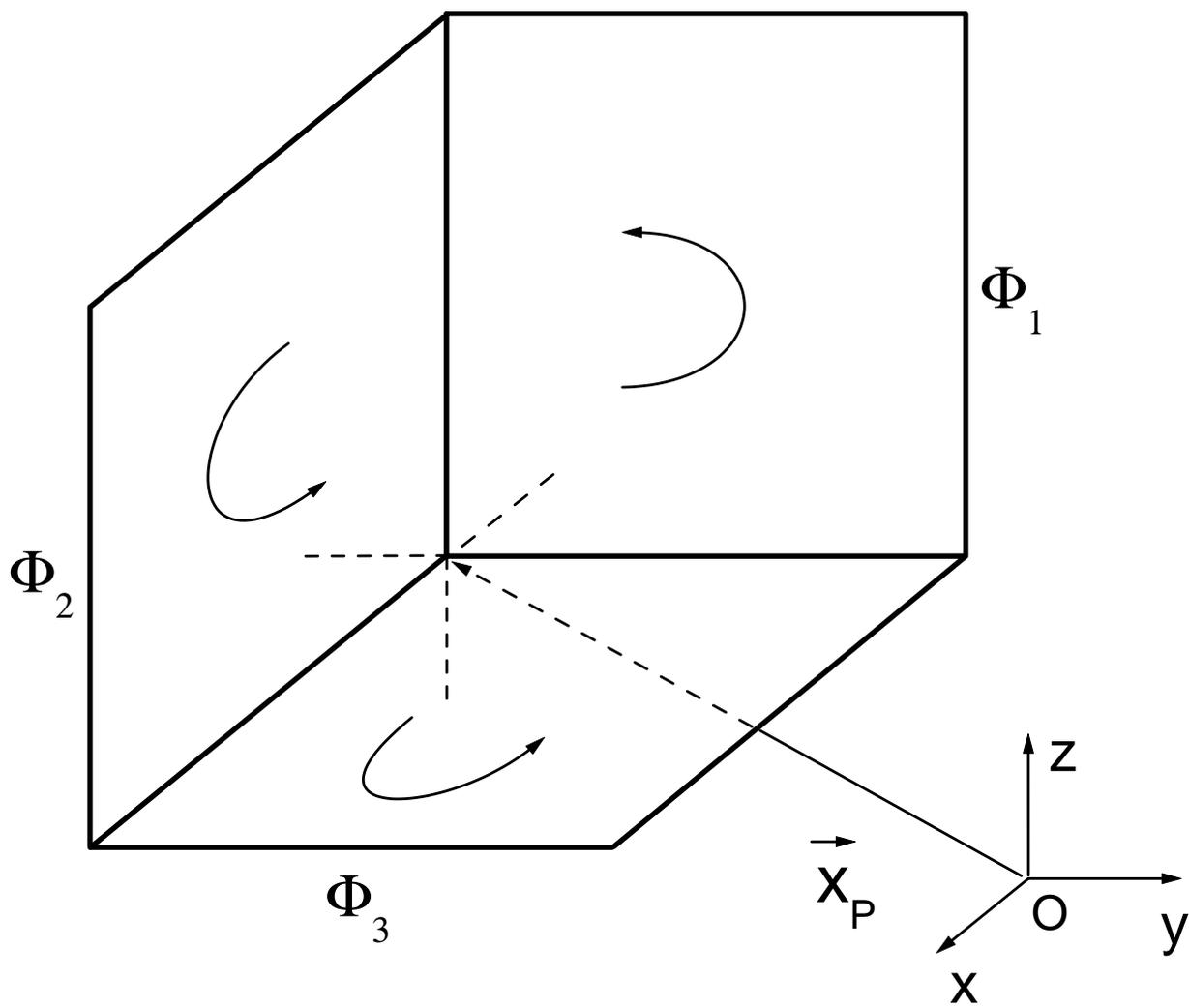
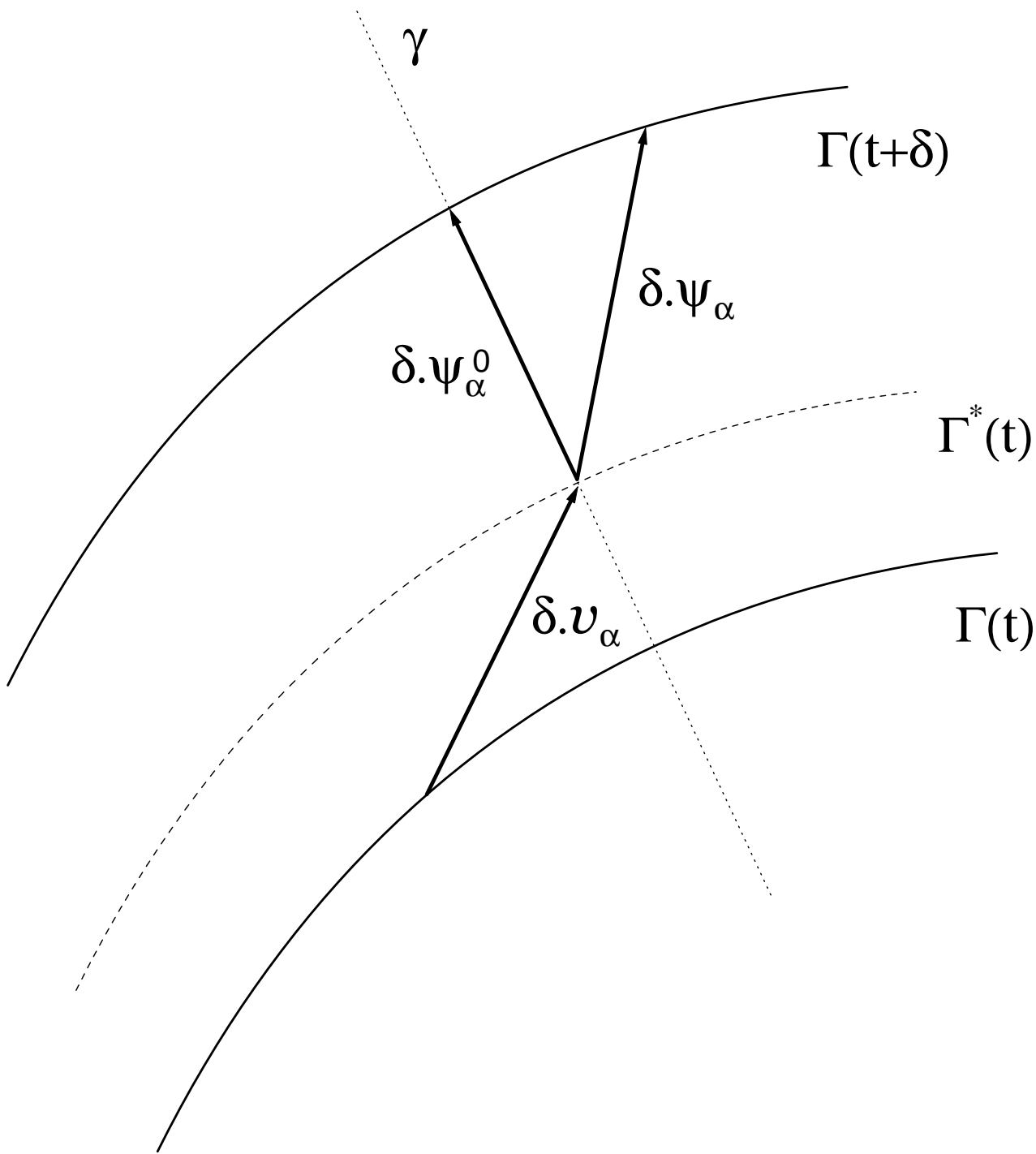
\end{document}